\documentclass[a4paper,11pt]{article}
\usepackage{pos}

\title{First measurement of $\rm \Omega_c^0$ production in pp collisions at $\sqrt{s}=13$ TeV with ALICE}

\manuallySeparateAuthors
\author*[a,b]{Jianhui Zhu}
\author{ for the ALICE Collaboration}

\affiliation[a]{Institute of Particle Physics, Central China Normal University,\\
  No.152 Luoyu Road, Wuhan, China}

\affiliation[b]{GSI Helmholtz Centre for Heavy Ion Research,\\
Planckstraße 1, Darmstadt, Germany}

\emailAdd{jianhui.zhu@cern.ch}

\abstract{Recent measurements of charm-baryon production in proton--proton (pp) collisions at midrapidity by the ALICE collaboration showed that baryon-to-meson yield ratios are significantly higher than those measured in $\rm e^+e^-$ collisions. The charm baryon-to-meson and charm baryon-to-baryon yield ratios provide unique information on hadronization mechanisms since the contributions from parton distribution function and parton--parton scattering terms cancel in the ratios. In this contribution, the first measurement of $\rm \Omega_{c}^{0}$ production via the hadronic decay channel $\rm \Omega_{c}^{0} \rightarrow \Omega^{-}\pi^{+}$ (and its charge conjugate) in $2<p_{\rm T}<12$~GeV/$c$ performed with the ALICE detector at midrapidity in pp collisions at $\sqrt{s}=$ 13 TeV is presented. The $\rm \Omega_c^0/D^0$ and $\rm \Omega_c^0/\Xi_c^0$ ratios multiplied by the decay branching ratio $\rm BR(\Omega_{c}^{0} \rightarrow \Omega^{-}\pi^+)$, which is not experimentally measured, are compared to MC generators with fragmentation fractions based on $\rm e^+e^-$ measurements and models including hadronization of charm quark via coalescence.}

\FullConference{%
 
 The Ninth Annual Conference on Large Hadron Collider Physics - LHCP2021\\
 7-12 June 2021\\
 Online
}

\begin{document}
\maketitle

\section{Introduction}
The production of heavy-flavor hadrons can be described by the factorization approach, which is a convolution of three independent terms: the parton distribution functions of the incoming partons, the cross sections of the partonic scatterings producing the heavy quarks, and the fragmentation functions that parametrise the non-perturbative evolution of a heavy quark into a given species of heavy-flavor hadron. Current perturbative quantum chromodynamics (pQCD) models based on this approach use fragmentation functions tuned on $\rm e^+e^-$ and electron--proton measurements and assume them to be universal across different colliding energy and systems. The heavy-flavor baryon-to-meson and baryon-to-baryon yield ratios are ideal observables related to the hadronization mechanism since the contributions from parton distribution function and parton--parton scattering terms cancel in the ratios. Recent measurements of charm-baryon production at midrapidity by the ALICE collaboration \cite{ALICE:2020wla, ALICE:2020wfu, ALICE:2021rzj, ALICE:2021psx, ALICE:2021bli, ALICE:2021dhb} show that the baryon-to-meson yield ratios are significantly higher than those in $\rm e^+e^-$ collisions, suggesting that the charm fragmentation is not universal. Thus, measurements of charm-baryon production are crucial to study the charm-quark hadronization in proton--proton (pp) collisions. Several mechanisms, such as color reconnection (CR) beyond the leading-color approximation \cite{Christiansen:2015yqa}, coalescence \cite{Song:2018tpv, Minissale:2020bif}, and feed-down from a largely augmented set of higher mass charm-baryon states beyond the current listings of the Particle Data Group (PDG)  \cite{ParticleDataGroup:2020ssz, He:2019tik}, have been proposed to explain this enhancement. The measurement of $\rm \Omega_c^0$ performed with the ALICE detector will be used to further constrain these mechanisms.

\section{Experimental apparatus and data analysis}

\begin{figure}[!ht]
\centering
\includegraphics[width=0.49\textwidth]{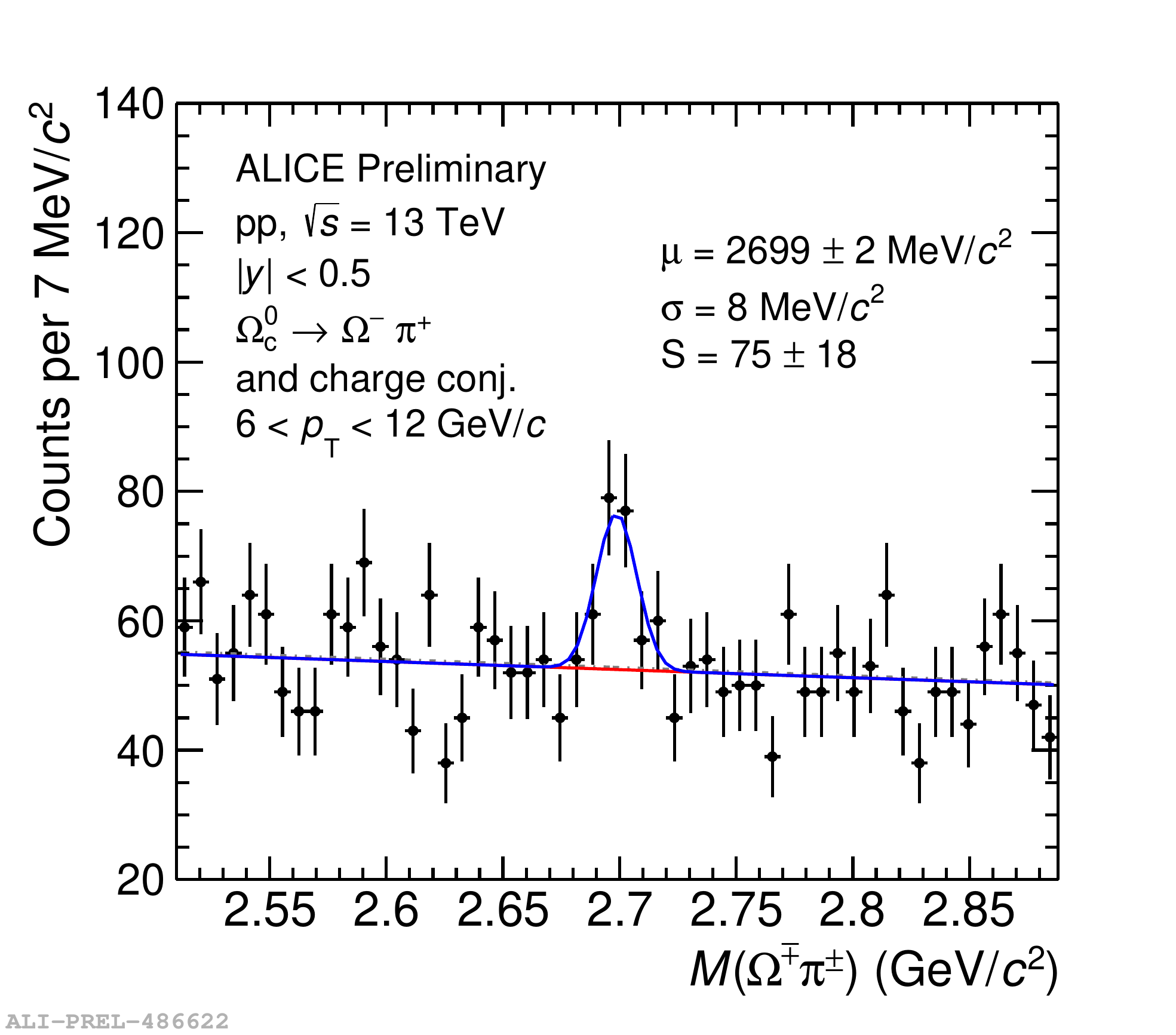}
\includegraphics[width=0.49\textwidth]{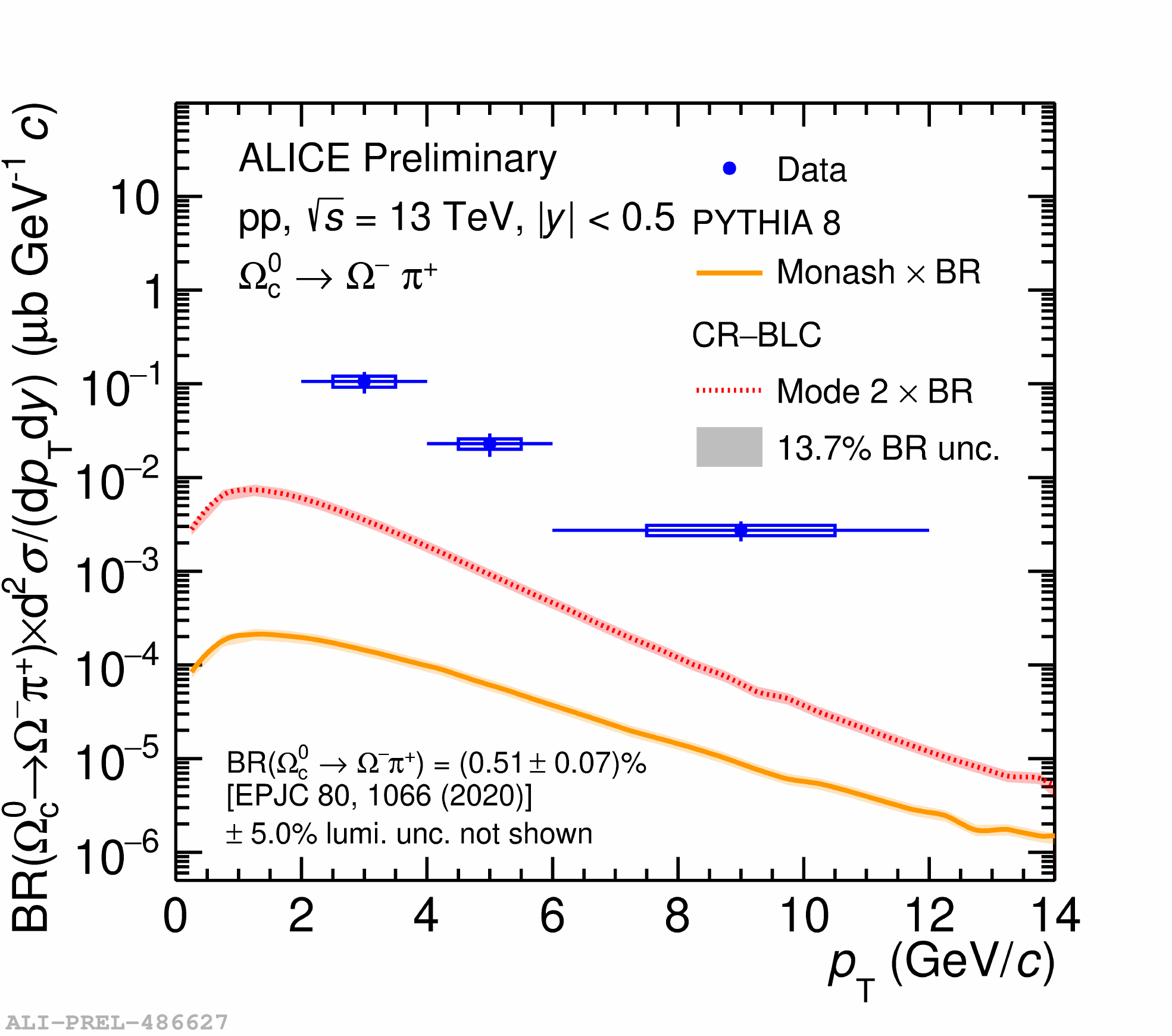}
\caption{Left: Invariant-mass distribution of $\rm \Omega_c^0\rightarrow\Omega^-\pi^+$ candidates and their charge conjugates in the interval $6<p_{\rm T}<12$ GeV/$c$. The blue line shows the total fit function and the red line represents the combinatorial background. Right: Inclusive $\rm \Omega_c^0$-baryon $p_{\rm T}$-differential production cross section multiplied by the $\rm BR(\Omega_c^0\rightarrow\Omega^-\pi^+)$ as a function of $p_{\rm T}$ for $|y|<0.5$ in pp collisions at $\sqrt{s}=13$ TeV. The error bars and empty boxes represent the statistical and systematic uncertainties, respectively. The measurement is compared with PYTHIA 8 with Monash tune \cite{Skands:2014pea} and with CR beyond the leading-color approximation \cite{Christiansen:2015yqa}, which are multiplied by a theoretical $\rm BR(\Omega_c^0\rightarrow\Omega^-\pi^+)=(0.51\pm0.07)\%$ \cite{Hsiao:2020gtc}.}
\label{fig_1}
\end{figure}

The ALICE apparatus is equipped with various detectors for triggering, tracking and particle identification (PID) \cite{Aamodt:2008zz}. The main detectors used for this measurement are the Inner Tracking System (ITS), the Time Projection Chamber (TPC) and the Time-Of-Flight detector (TOF), which are located in the central barrel at midrapidity ($|\eta|<0.9$) embedded in a solenoidal magnet that provides a $\rm B=0.5$~T field parallel to the beam direction. The ITS is used for tracking and vertex reconstruction. The TPC is the main tracking detector in the central barrel and is also used for particle identification via specific energy loss (${\rm d}E/{\rm d}x$) measurements. The TOF provides complementary PID informations via measurements of the time of flight. The $\rm \Omega_c^0$ baryon is reconstructed via the hadronic decay channel $\rm \Omega^-\pi^+$, together with their charge conjugates, using the Kalman-Filter vertexing algorithm \cite{KalmanFilter} in the transverse momentum ($p_{\rm T}$) interval $2<p_{\rm T}<12$ GeV/$c$. The decay topological selections on the daughter tracks of $\rm \Omega_c^0$ baryon with machine learning techniques are implemented to reject background. The signal extraction is performed by the invariant mass analysis as shown in the left panel of Fig.~\ref{fig_1}. Then, the yields are corrected for acceptance and efficiency and are converted into the cross section multiplied by the decay branching ratio (BR).

\section{$p_{\rm T}$-differential production of inclusive $\rm \Omega_c^0$}

The $\rm BR(\Omega_c^0\rightarrow\Omega^-\pi^+)$ is not measured, hence the $\rm BR(\Omega_c^0\rightarrow\Omega^-\pi^+)$ $\times {\rm d}^2\sigma/({\rm d}p_{\rm T}{\rm d}y)$ of the $\rm \Omega_c^0$ baryon as a function of $p_{\rm T}$ for $|y|<0.5$ in pp collisions at $\sqrt{s}=13$ TeV is shown in the right panel of Fig.~\ref{fig_1}. The feed-down contribution is not subtracted. PYTHIA 8 with Monash tune \cite{Skands:2014pea} and with CR beyond the leading-color approximation \cite{Christiansen:2015yqa}, which are multiplied by the BR from a theoretical calculation $\rm BR(\Omega_c^0\rightarrow\Omega^-\pi^+)=(0.51\pm0.07)\%$ \cite{Hsiao:2020gtc}, largely underestimate the measurement.

\section{Baryon-to-meson and baryon-to-baryon yield ratios}

\begin{figure}[!ht]
\centering
\includegraphics[width=0.49\textwidth]{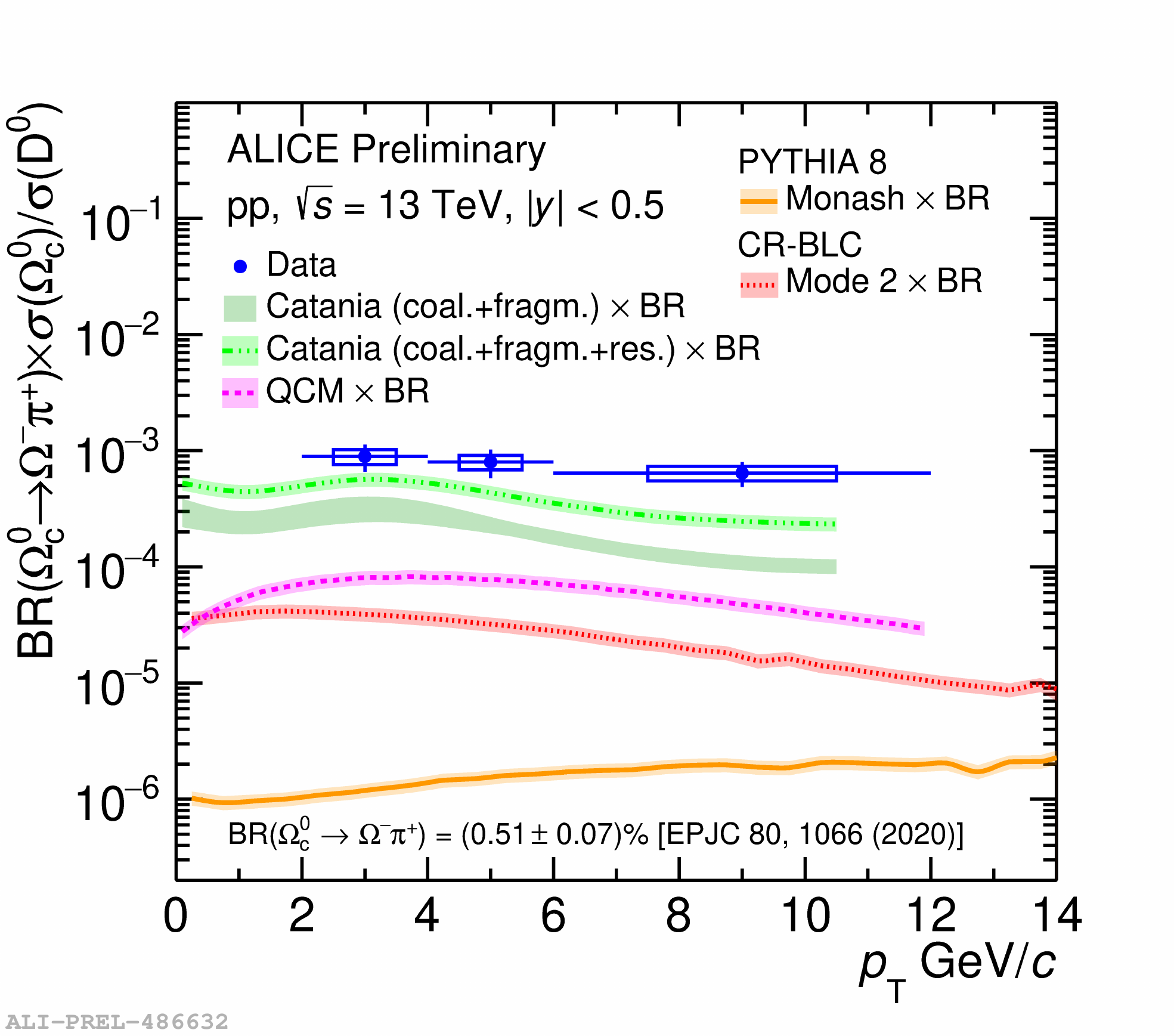}
\includegraphics[width=0.49\textwidth]{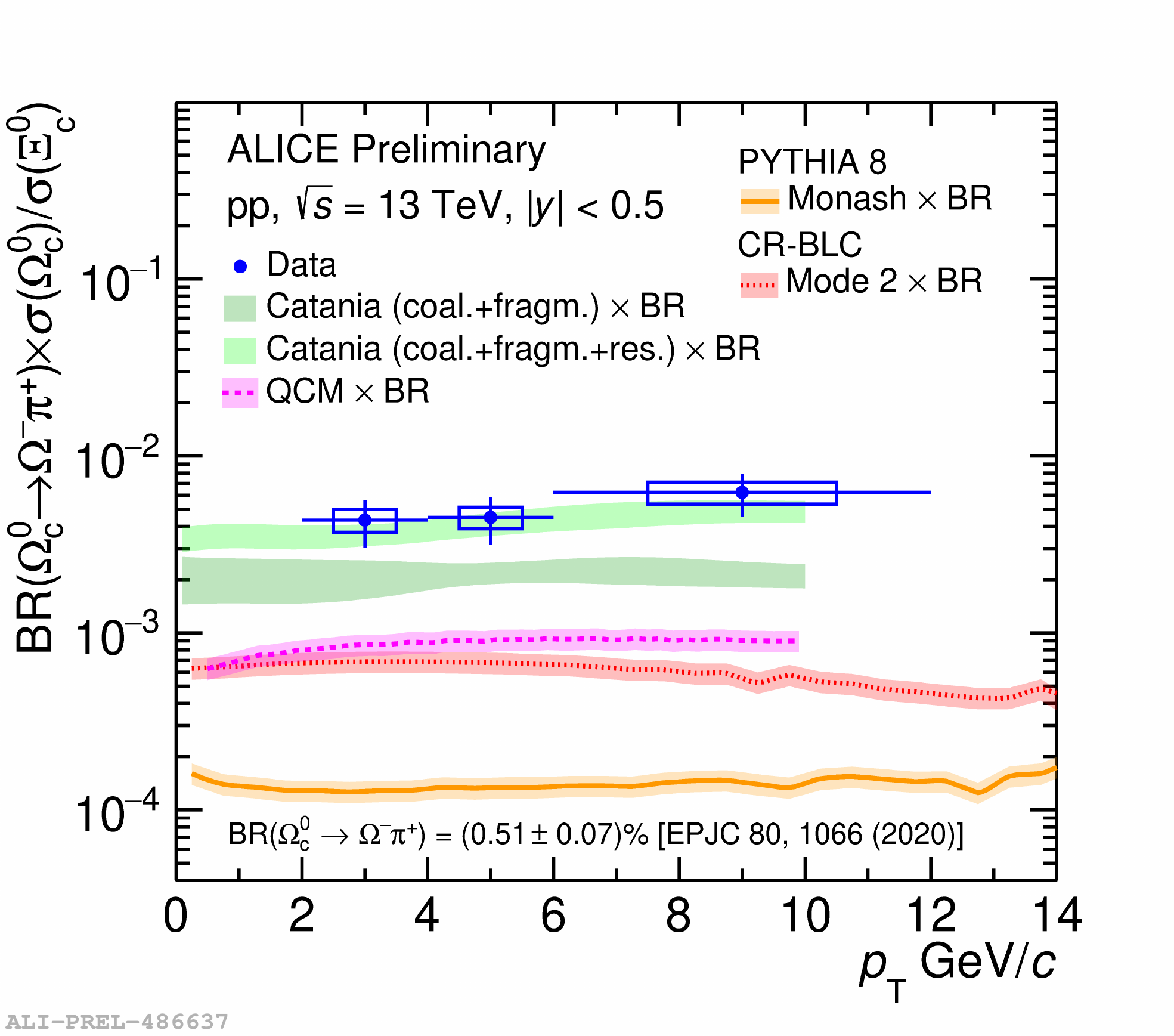}
\caption{The ${\rm BR(\Omega_c^0\rightarrow\Omega^-\pi^+)} \times {\rm d}^2\sigma/({\rm d}p_{\rm T}{\rm d}y)$ of inclusive $\rm \Omega_c^0$ baryons over ${\rm d}^2\sigma/({\rm d}p_{\rm T}{\rm d}y)$ of prompt $\rm D^0$ mesons (left) and over ${\rm d}^2\sigma/({\rm d}p_{\rm T}{\rm d}y)$ of prompt $\rm \Xi_c^0$ baryons (right) as a function of $p_{\rm T}$ in pp collisions at $\sqrt{s}=13$ TeV. The error bars and empty boxes represent the statistical and systematic uncertainties, respectively. The measurement is compared with model calculations, which are multiplied by a theoretical $\rm BR(\Omega_c^0\rightarrow\Omega^-\pi^+)=(0.51\pm0.07)\%$ \cite{Hsiao:2020gtc}.}
\label{fig_2}
\end{figure}

The ratios of the $p_{\rm T}$-differential $\rm BR(\Omega_c^0\rightarrow\Omega^-\pi^+)\times cross\ section$ of $\rm \Omega_c^0$ baryons to the cross section of $\rm D^0$ mesons and to the cross section of $\rm \Xi_c^0$ baryons are shown in the left and right panel of Fig.~\ref{fig_2}, respectively. All the models are scaled by the theoretical $\rm BR(\Omega_c^0\rightarrow\Omega^-\pi^+)=(0.51\pm0.07)\%$ \cite{Hsiao:2020gtc}. PYTHIA 8 with Monash tune \cite{Skands:2014pea} and with CR beyond the leading-color approximation \cite{Christiansen:2015yqa} largely underestimate both baryon-to-meson and baryon-to-baryon yield ratios. The measured ratios are also compared with models that include hadronization via coalescence.  In the quark combination mechanism (QCM) \cite{Song:2018tpv}, the charm quark can pick up a co-moving light antiquark or two co-moving quarks to form a single-charm meson or baryon. The model does not describe these two ratios. The Catania model \cite{Minissale:2020bif} implements charm-quark hadronization via both coalescence and fragmentation, and it is the model that gets closer to the measured ratios over the full $p_{\rm T}$ interval, especially when additional resonance states are considered as given by the PDG \cite{ParticleDataGroup:2020ssz}.

\section{Summary}
The first LHC measurement of the inclusive $p_{\rm T}$-differential production of the charm-strange baryon $\rm \Omega_c^0$ multiplied by the branching ratio into $\rm \Omega^-\pi^+$ at midrapidity in pp collisions at $\sqrt{s}=13$~TeV is reported. The ratios of this measurement to the production cross section of the $\rm D^0$ meson and to that of the $\rm \Xi_c^0$ baryon are largely underestimated by theoretical calculations, except the Catania model, which considers coalescence, fragmentation and additional resonance states in pp collisions.

\section{Acknowledgments}
This work was supported by the National Natural Science Foundation of China (NSFC) (No. 12105109) and the international postdoctoral exchange fellowship program of Helmholtz Association and the Office of China Postdoc Council (No. 20181016).

\bibliographystyle{utphys}
\bibliography{references}

\end{document}